\documentclass[11pt]{article}
\usepackage{epsfig}
\usepackage{amsfonts}
\usepackage{amsmath}
\usepackage{amssymb}
\usepackage{dsfont}
\usepackage{hyperref}
\usepackage{color}
\usepackage{cite}

\newcommand{\sq}{^{2}}

\newcommand{\dslash}[1]{#1 \llap{/\kern-0.5pt}}
\newcommand{\Dslash}[1]{#1 \llap{/\kern+1.5pt}}
\newcommand{\DDslash}[1]{#1 \llap{/\kern+2.3pt}}
\newcommand{\dslashh}[1]{#1 \llap{/\kern+1pt}}

\newcommand{\Ex}[1]{\cdot 10^{#1}}
\newcommand{\bea}{\begin{eqnarray}}
\newcommand{\eea}{\end{eqnarray}}
\newcommand{\be}{\begin{equation}}
\newcommand{\ee}{\end{equation}}
\newcommand{\bma}{\begin{pmatrix}}
\newcommand{\ema}{\end{pmatrix}}
\newcommand{\nn}{\nonumber}

\newcommand{\NLDBD}{$0 \nu \beta \beta$}

\begin{document}
\begin{titlepage}

\begin{flushright}
LA-UR-17-20043
\end{flushright}

\vspace{2.0cm}

\begin{center}
{\LARGE  \bf 
Neutrinoless 
double beta decay
and  chiral $SU(3)$ 
}
\vspace{2.4cm}

{\large \bf  V. Cirigliano$^a$, W. Dekens$^{a,b}$,  M. Graesser$^a$, and E. Mereghetti$^a$ } 
\vspace{0.5cm}

\vspace{0.25cm}

{\large 
$^a$ 
{\it Theoretical Division, Los Alamos National Laboratory,
Los Alamos, NM 87545, USA}}

\vspace{0.25cm}
{\large 
$^b$ 
{\it 
New Mexico Consortium, Los Alamos Research Park, Los Alamos, NM 87544, USA
}}

\end{center}

\vspace{1.5cm}

\begin{abstract}

TeV-scale lepton number violation can affect neutrinoless double beta decay  
 through dimension-9  $\Delta L= \Delta I =  2$  operators involving two electrons and four quarks. 
Since the dominant effects within a nucleus are expected to arise from pion exchange,  
the $ \pi^- \to \pi^+   e e$ matrix elements of the dimension-9 operators are a key hadronic input.  
In this letter we  provide estimates for the $\pi^- \to \pi^+ $ matrix elements 
of  all  Lorentz scalar $\Delta I = 2$  four-quark operators relevant to the study of 
TeV-scale lepton number violation.    The  analysis is based on   
chiral $SU(3)$ symmetry, which relates the $\pi^- \to \pi^+$  matrix elements of the $\Delta I = 2$ operators 
to the  $K^0 \to \bar{K}^0$  and  $K \to \pi \pi$   matrix elements of their $\Delta S = 2$ and $\Delta S = 1$ chiral partners,  for which lattice  QCD input is available. 
The inclusion of  next-to-leading order chiral  loop corrections to all  symmetry relations used in the analysis  
makes  our  results robust at the 30\% level or better, depending on the operator.

\end{abstract}

\vfill
\end{titlepage}

{\bf Introduction} --   Neutrinoless double beta decay  (\NLDBD) is a rare nuclear process in which two neutrons inside a nucleus convert into two protons with emission of two electrons and no neutrinos, thus changing the number of leptons by two units. Since lepton number is conserved in the Standard Model (SM) at the classical level, observation of \NLDBD \  would be direct evidence of new physics, with far reaching implications: it would demonstrate that neutrinos are Majorana fermions~\cite{Schechter:1981bd},  shed light on the mechanism of neutrino mass generation, and probe lepton number violation (LNV), a key ingredient needed to generate the matter-antimatter asymmetry in the universe via  ``leptogenesis''~\cite{Davidson:2008bu}. 
The current experimental  limits on the half-lives are already 
impressive~\cite{KamLAND-Zen:2016pfg, Alfonso:2015wka,Albert:2014awa,Agostini:2013mzu,Gando:2012zm,Elliott:2016ble,Andringa:2015tza},
at the level  of  $T_{1/2} >  2.1\times10^{25}$~y  for $^{76}$Ge~\cite{Agostini:2013mzu}
and  $T_{1/2} >  1.07\times10^{26}$~y  for $^{136}$Xe~\cite{KamLAND-Zen:2016pfg}, 
with  next generation ton-scale experiments aiming at a sensitivity of  $T_{1/2} \sim 10^{27-28}$~y.

By itself, the observation of \NLDBD \   would not  immediately point to the underlying physical origin of  LNV. 
While \NLDBD  \ searches are  commonly interpreted in terms of the exchange of a light Majorana neutrino,  
other new physics mechanisms  deserve careful evaluation.   
In an effective theory approach to new physics, the light Majorana neutrino exchange  dominates 
whenever the scale of lepton number violation, $\Lambda_{\rm LNV}$, is very high compared to the electroweak scale:  
as long as $\Lambda_{\rm LNV} \gg$~TeV,  the only low-energy manifestation of this new physics is a Majorana mass 
for light neutrinos, encoded in a  single gauge-invariant dimension-5 operator~\cite{Weinberg:1979sa}.
However, as  $\Lambda_{\rm LNV}$ is lowered,  new contributions to \NLDBD  \ are possible, 
which typically involve the exchange of new TeV-mass Majorana fermions, for example R-handed neutrinos in left-right symmetric models  
or neutralinos in certain supersymmetric models (for recent  reviews see Refs.~\cite{Rodejohann:2011mu,deGouvea:2013zba,DellOro:2016tmg,Engel:2016xgb}).
At low energy,  the effects of this TeV scale LNV dynamics can be encoded in a set of local  dimension-9 operators (involving two leptons and four quarks)  
that change lepton number by two units.   The operators  have been classified both 
according to $SU(3)_C \times U(1)_{EM}$  gauge invariance ~\cite{Pas:2000vn,Prezeau:2003xn}, 
directly relevant at low-energy,  and  $SU(3)_C \times SU(2)_W \times U(1)_Y$  gauge invariance~\cite{Graesser:2016bpz}, 
important to connect  \NLDBD \  to possible LNV signals  at the Large Hadron Collider.

\begin{figure}
\center
\includegraphics[width=0.8\textwidth]{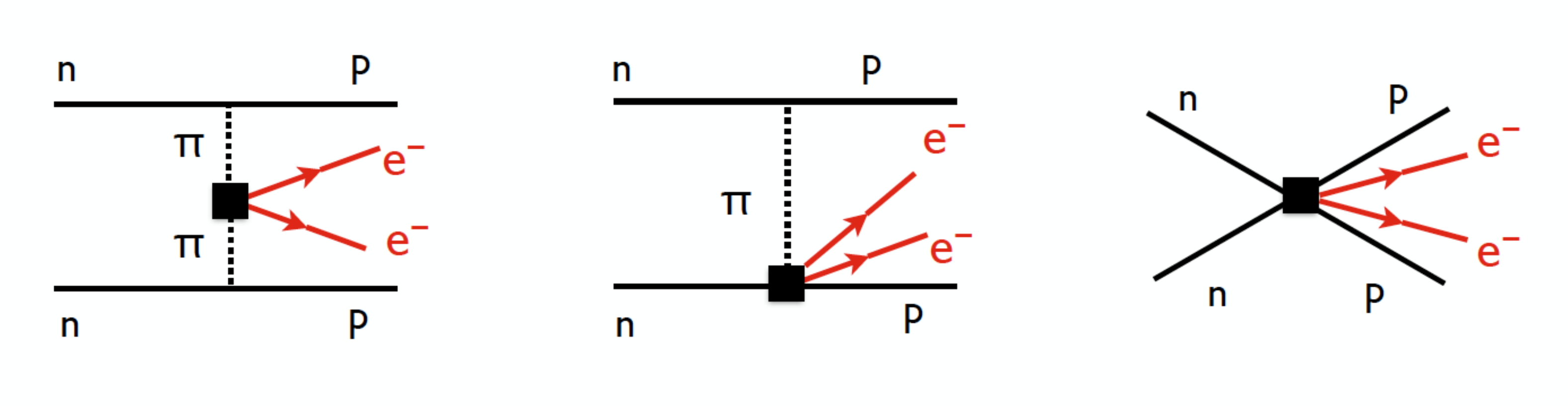}
\caption{Feynman diagrams representing the insertion of dimension-9 operators of Eq.~\eqref{eq:Leff1} -- denoted by a black square --  at the hadronic level. 
In this paper we  study  the $\pi^- \to \pi^+  e e$  vertex appearing in the leftmost diagram, which is enhanced  in the chiral power counting.  
\label{fig:fig1}}
\end{figure}

To interpret positive or null \NLDBD \ results in the context of TeV-scale LNV dynamics, it is essential  to 
quantify the hadronic and nuclear matrix elements involving the $\Delta L=2$ dimension-9 operators. 
This is conveniently tackled by first matching the dimension-9 quark level operators onto appropriate 
operators at the pion-nucleon level, and subsequently computing the nuclear matrix elements. 
 As illustrated  in Fig.~\ref{fig:fig1}, the dimension-9 operators induce  a variety of effective vertices at the pion-nucleon level.
 The use of chiral power counting has led to the identification of  the two-pion exchange in Fig.~\ref{fig:fig1}
as the dominant contribution~\cite{Faessler:1996ph,Prezeau:2003xn}.
It is therefore  very important to estimate as accurately as possible the $ \pi^- \pi^- \to e e$ matrix elements of the dimension-9 operators. 
Current knowledge of these matrix elements is based on vacuum saturation or naive dimensional analysis, 
with the exception of two operators  for which  chiral symmetry was used to 
relate the two-pion matrix elements to the $K^\pm \to \pi^\pm \pi^0$ amplitude~\cite{Savage:1998yh}. 

In this paper we  generalize  the chiral symmetry analysis  to all  Lorentz scalar $\Delta I = 2$  four-quark operators, $O_{1,...,5}$ defined in 
Eq.~\eqref{eq:basis} below. 
We provide estimates for the   $\pi^- \to \pi^+$  matrix elements of $O_{2,...,5}$  
by relating them to the  $K^0 \to \bar{K}^0$ matrix elements of their $\Delta S = 2$ chiral partners, 
which have been computed by several lattice QCD groups~\cite{Carrasco:2015pra,Bertone:2012cu,Jang:2015sla,Boyle:2012qb,Garron:2016mva}. 
By including  the  leading chiral loop corrections,
we  are able to estimate  the uncertainty on the symmetry relations, finding that it does not exceed 30\%.

{\bf Operator basis and chiral transformation properties} -- 
At the  hadronic scale, short-distance contributions to $0\nu \beta \beta$ can 
be parameterized by a number of dimension-9 operators~\cite{Prezeau:2003xn,Gonzalez:2015ady,Graesser:2016bpz}
\be
{\cal L}_{\rm eff} =  \frac{1}{\Lambda_{\rm LNV}^5}  \left[ \sum_{i = {\rm scalar} }   \,  \left( c_{i, S}  \, \bar{e} e^c  +  c_{i,S}' \, \bar{e} \gamma_5 e^c \right) \, O_{i} 
\ + \ 
\bar{e}  \gamma_\mu \gamma_5 e^c \ \sum_{i = {\rm  vector} }   \,  c_{i, V}   \, {O}_{i}^\mu 
\right] ~, 
\label{eq:Leff1}
\ee
where $O_i$ and $O_j^\mu$ denote scalar and vector four-quark operators, respectively. 
In this letter we focus on the scalar operators and in order to discuss their properties 
under the chiral $SU(3)_L \times SU(3)_R$ group we chose to work in the following basis~\footnote{
This is consistent with the bases  used  in Refs.~\cite{Gabbiani:1996hi} and \cite{Buras:2000if} 
for the  $\Delta S = 2$ effective Hamiltonian beyond the Standard Model. 
Compared to the basis presented in Ref.~\cite{Graesser:2016bpz}, we are able to eliminate 
the operator involving tensor densities $\sigma_{\mu \nu} \otimes \sigma^{\mu \nu}$. }
\begin{subequations}
\bea
O_ 1  &=&  \bar{q}_L^\alpha  \gamma_\mu \tau^+ q_L^\alpha \  \  \bar{q}_L^\beta  \gamma^\mu \tau^+ q_L^\beta    
\\
O_ 2  &=&  \bar{q}_R^\alpha  \tau^+ q_L^\alpha \ \   \bar{q}_R^\beta  \tau^+ q_L^\beta    
\\
O_ 3  &=&  \bar{q}_R^\alpha  \tau^+ q_L^\beta \  \ \bar{q}_R^\beta  \tau^+ q_L^\alpha    
\\
O_ 4  &=&  \bar{q}_L^\alpha  \gamma_\mu \tau^+ q_L^\alpha \ \  \bar{q}_R^\beta  \gamma^\mu \tau^+ q_R^\beta    
\\
O_ 5  &=&  \bar{q}_L^\alpha  \gamma_\mu \tau^+ q_L^\beta \ \  \bar{q}_R^\beta  \gamma^\mu \tau^+ q_R^\alpha    ~,
\eea
\label{eq:basis}%
\end{subequations}
where $ q^T = (u, d, s)$, $q_{L,R} =  (1/2)( 1 \mp \gamma_5)  q$,  $\alpha$, $\beta$ denote color indices, and $\tau^+ =  T^1 + i T^2$ in terms of the $SU(3)$ generators $T^a$. 
Three additional operators $O_{1,2,3}'$  are obtained from  $O_{1,2,3}$ by the interchange $L \leftrightarrow R$ everywhere. 
Parity invariance of QCD implies  $\langle \pi^+ | O_{1,2,3}' | \pi^- \rangle  = \langle \pi^+ | O_{1,2,3}| \pi^- \rangle$. 

The operators $O_i$ belong to irreducible representations of the  
chiral symmetry group  $SU(3)_L \times SU(3)_R$   ($q_{L,R} \to  U_{L,R} q_{L,R}$ with $U_{L,R} \in  SU(3)_{L,R}$).
$O_1$ transforms as ${\bf 27}_L \times {\bf 1}_R$, 
$O_{2,3}$  transform as ${\bf 6}_L \times {\bf \bar{6}}_R$,  and finally  
$O_{4,5}$  transform as ${\bf 8}_L \times {\bf 8}_R$.
The transformation properties of $O_1$ were exploited in Ref.~\cite{Savage:1998yh} to  relate the 
matrix element of $ \langle \pi ^+ | O_1 | \pi^- \rangle $  to the $\Delta I = 3/2$ $K^+ \to \pi^- \pi^0$ amplitude. 
Here we exploit the transformation properties of $O_{2,3,4,5}$ to relate their two-pion matrix elements 
to the matrix elements of their chiral partners between a $K^0$ and a $\bar{K}^0$ meson, which 
have been computed with lattice QCD by several groups~\cite{Carrasco:2015pra,Bertone:2012cu,Jang:2015sla,Boyle:2012qb,Garron:2016mva}. 
Strictly speaking the symmetry relation  is valid only to leading order in the chiral expansion, and is expected to receive $O(30\%)$ corrections. 
To make our analysis more robust,  we also estimate the size of next-to-leading order (NLO)  quark-mass  corrections by computing the leading chiral loops.

\begin{figure}
\center
\includegraphics[width=0.8\textwidth]{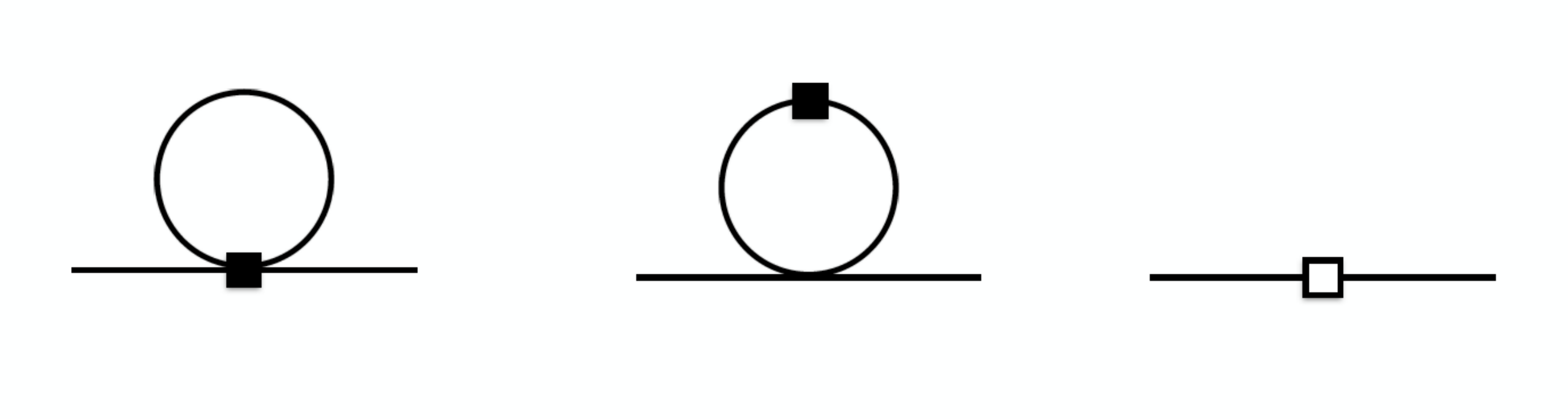}
\vspace{-0.5cm}
\caption{Feynman diagrams  contributing to the NLO corrections to ${\cal M}^{\pi \pi}$ and  
${\cal M}^{K \bar K}$.  Solid lines represent $\pi, K, \eta$. The black squares represent an insertion 
of the lowest-order chiral Lagrangian (see Eq.~\eqref{eq:chiral1}), while 
the open square represents an insertion from the NLO effective Lagrangian. 
\label{fig:fig2}}
\end{figure}

{\bf  Determination of $\mathbf{\langle \pi^+ | O_{2,3,4,5}| \pi^- \rangle}$} --  
The argument proceeds as follows. 
 $O_{2,3}$  and $O_{4,5}$  can be written as linear combinations of operators 
transforming according to  the ${\bf 6}_L \times {\bf \bar{6}}_R$ and ${\bf 8}_L \times {\bf 8}_R$ representations of the chiral group, respectively. 
These operators in turn admit a unique  hadronic realization to leading order in the chiral expansion
\begin{subequations}
\bea
O_{6 \times \bar 6}^{a,b} =  \bar{q}_R T^a  q_L \  \  \bar{q}_R  T^b  q_L     & \to &   g_{6 \times \bar{6}}   \  
\frac{F_0^4}{4}  \, 
 {\rm Tr} \left( T^a U  T^b U \right) 
 \\
O_{8 \time 8}^{a,b} =  \bar{q}_L  T^a  \gamma_\mu  q_L \ \   \bar{q}_R T^b  \gamma^\mu  q_R    & \to &   g_{8 \times 8}   \ 
\frac{F_0^4}{4}  \, 
  {\rm Tr} \left( T^a U  T^b U^\dagger \right) ~, 
\eea
\label{eq:chiral1}%
\end{subequations}
where  the trace is over flavor and     $U$ is the usual matrix of  pseudo-Nambu-Goldstone boson fields 
transforming as $U \to  U_L  \, U \, U_R^\dagger$ under $SU(3)_L \times SU(3)_R$, 
\begin{equation}\label{eq:2.3}
U   = \exp\left(\frac{\sqrt{2} i \pi}{F_0}\right), \qquad
\pi  =  \left( \begin{array}{c c c} 
\frac{\pi_3}{\sqrt{2}} + \frac{\pi_8}{\sqrt{6}}  & \pi^+ 						& K^+ \\
\pi^-						& - \frac{\pi_3}{\sqrt{2}} + \frac{\pi_8}{\sqrt{6}} 	& K^0 \\
K^-						& \bar{K}^0					& - \frac{2}{\sqrt{6}}\pi_8 
\end{array} \right)\, ,
\end{equation}
and $F_0$  is the pseudoscalar decay constant in the chiral limit (in our normalization $F_\pi \simeq 92.4$~MeV). 
The non-perturbative dynamics is encoded in  the  low-energy constants $g_{6 \times \bar{6}}$,  $g_{8 \times 8}$,  and  
for each representation there are two independent constants, corresponding to different color contractions (e.g. $O_2$ and $O_3$).

The operators in Eqs.~\eqref{eq:basis} are obtained by setting $T^a = T^b \to  T^1 + i T^2$ in Eq.~\eqref{eq:chiral1}. 
The same representations, however,  contain $\Delta S = 2$ operators that contribute to $K^0$-$\bar{K}^0$ mixing  in extensions of the Standard Model 
($T^a = T^b \to  T^6 -i T^7$ in Eq.~\eqref{eq:chiral1}). 
The relevant  $K^0$-$\bar{K}^0$ matrix elements have been  computed in lattice QCD, thus providing the couplings $g_{6 \times \bar{6}}$ and $g_{8 \times 8}$ 
to leading order in the chiral $SU(3)$ expansion. Note  that  the  $g_{8 \times 8}$  couplings can be  independently extracted  through their 
contributions to $K^0 \to (\pi \pi)_{I=2}$ amplitudes via the $\Delta S = 1$  electroweak penguin operators, that  transform as  ${\bf 8}_L \times {\bf 8}_R$.

We first focus on the relation to $K^0$-$\bar{K}^0$  mixing.  A straightforward calculation based on the leading chiral realization of 
Eq.~\eqref{eq:chiral1} leads to:
\begin{subequations}
\bea
{\cal M}^{\pi \pi}_{6 \times \bar 6} \equiv 
\langle \pi^+ |  O_{6 \times \bar 6}^{1+ i2, 1+i2}  | \pi^- \rangle &=&    \langle \bar{K}^0  |  O_{6 \times \bar 6}^{6- i7, 6-i 7}  | K^0 \rangle 
\equiv {\cal M}^{K \bar K}_{6 \times \bar 6}  
\\
{\cal M}^{\pi \pi}_{8 \times 8} \equiv 
\langle \pi^+ |  O_{8  \times 8}^{1+ i2, 1+i2}  | \pi^- \rangle  &=&    \langle \bar{K}^0  |  O_{8 \times 8}^{6- i7, 6-i 7}  | K^0 \rangle 
\equiv {\cal M}^{K \bar K}_{8 \times 8}  
~.
\eea
\label{eq:relation1}%
\end{subequations}
NLO  chiral corrections arising from the one-loop diagrams of Fig.~\ref{fig:fig2} 
and $O(p^2)$ counterterms   could alter the above relation (see for example  Refs.~\cite{Donoghue:1982cq} and \cite{Bijnens:1984ec} for the 
analogous discussion of  $K^0$-$\bar{K}^0$ mixing and $K^\pm \to \pi^\pm \pi^0$ amplitudes in the Standard Model). 
For the relations of interest here,  we find   to NLO at zero momentum transfer (defining $L_{\pi,K,\eta} \equiv \log  \mu_\chi^2 / m_{\pi, K, \eta}^2$)
\begin{eqnarray}
{\cal M}^{K \bar K}_{8 \times 8}  
&=& g_{8 \times 8} \,  F_K^2 \left\{  1 + \frac{1}{(4 \pi F_0)^2} \left( m_K^2  \left(- 1 + 2 L_K\right) - \frac{m_\pi^2}{4} L_\pi - \frac{3}{4} m_\eta^2  L_\eta   
+ \delta^{K \bar K}_{8 \times 8}  \right)       \right\}
  \quad \ 
 \label{eq:88pipi1}
\\
{\cal M}^{\pi \pi}_{8 \times 8}  
  &=&  g_{8 \times 8} \,  F_\pi^2 \left\{  1 + \frac{1}{(4 \pi F_0)^2} \bigg(  m_\pi^2 \left( -1 +  L_\pi \right)   + \delta^{\pi \pi}_{8 \times 8}  \bigg) \right\}~.
 \label{eq:88pipi}
\end{eqnarray}
To identify  the finite parts of the loops we have followed  the modified $\overline{\rm MS}$ scheme commonly used 
chiral perturbation theory~\cite{Gasser:1984gg,Cirigliano:2003gt}. 
Moreover,   $\delta_{8 \times 8}^{K \bar K}$ and  $\delta_{8 \times 8}^{\pi \pi}$ denote linear combinations of $O(p^2)$ counterterms, 
that reabsorb the $\mu_\chi$ dependence of $L_{\pi, K, \eta}$
and contain additional finite corrections. 
Using the NLO  effective Lagrangian~\cite{Cirigliano:1999pv}, we find  
\be
 \delta^{K \bar K}_{8 \times 8}   =    a_{8\times8} \ m_K^2  +  b_{8 \times 8} \,  \left(m_K^2 + \frac{1}{2} m_\pi^2 \right)  \qquad \qquad 
 \delta^{\pi \pi}_{8 \times 8}   =    a_{8 \times 8}  \  m_\pi^2  +  b_{8 \times 8} \,  \left(m_K^2 + \frac{1}{2} m_\pi^2 \right)~,
\label{eq:LEC1}
\ee 
with $a_{8 \times 8}$ and $b_{8 \times 8}$  dimensionless  constants. 
Similarly,  for  the   ${\bf 6}_L \times {\bf  \bar{6}}_R$. representation we find
\begin{eqnarray}
{\cal M}^{K \bar K}_{6 \times \bar 6}  
\!\! &=& \!\!\!\!
 - g_{6  \times \bar 6} 
F_K^2 \left\{  1 + \frac{1}{(4 \pi F_0)^2} \left( m_K^2  \left(-1 + 2 L_K\right) + \frac{m_\pi^2}{4}  L_\pi - \frac{7}{12} m_\eta^2 L_\eta  
+ \delta^{K \bar K}_{6 \times \bar 6}   \right)   \right\} 
\ \ \ 
\\
{\cal M}^{\pi \pi}_{6 \times \bar 6}  
\!\! &=& \!\!\!\!
- g_{6  \times \bar 6} 
 F_\pi^2 \left\{  1 + \frac{1}{(4 \pi F_0)^2} \left( m_\pi^2 (-1 + L_\pi)  + \frac{2}{3} m_\eta^2 L_\eta 
 + \delta^{\pi \pi}_{6 \times \bar 6}    \right)     \right\}  ~,
\label{eq:mpipi6}
\end{eqnarray}
with counterterm contributions analogous to the ones in Eq.~\eqref{eq:LEC1}.
The loop corrections to ${\cal M}^{K \bar K}_{8\times8, 6 \times \bar 6}$ have been  calculated in Ref.~\cite{Becirevic:2004qd} 
and we agree with them. 
Eqs.~\eqref{eq:88pipi1}-\eqref{eq:mpipi6}  lead to the central result of our work, namely a  relation between 
${\cal M}^{\pi \pi}$ and ${\cal M}^{K \bar K}$ valid to NLO in the chiral expansion
\begin{subequations}
\bea
{\cal M}^{\pi \pi}_{8 \times 8}  &=&
{\cal M}^{K \bar K}_{8 \times 8} 
  \times    \frac{F_\pi^2}{F_K^2}   \times  \left( 1 +   \Delta_{8 \times 8} \right)  \ =  \   {\cal M}^{K \bar K}_{8 \times 8}  \times  R_{8 \times 8}
\\
{\cal M}^{\pi \pi}_{6 \times \bar 6}   &= &  
{\cal M}^{K \bar K}_{6 \times \bar 6}  \times 
\frac{F_\pi^2}{F_K^2}  \times 
\left (1 +  \Delta_{6 \times \bar 6} \right)  \  =  \
{\cal M}^{K \bar K}_{6 \times \bar 6}  \times 
 R_{6 \times \bar 6}~, 
\eea
\label{eq:relation2}
\end{subequations}
with 
\begin{subequations}
\bea
\Delta_{8 \times 8} &=&  \frac{1}{(4\pi F_0)^2} \left[ \frac{m_\pi^2}{4}(-4 + 5 L_\pi)  - m_K^2 (-1 + 2 L_K) +\frac{3}{4} m_\eta^2 L_\eta 
- a_{8 \times 8} \, \left(m_K^2 - m_\pi^2 \right)
\right] 
\\
\Delta_{6 \times \bar 6} &=& 
\frac{1}{(4\pi F_0)^2} \left[ - \frac{m_\pi^2}{4}(4 - 3 L_\pi)  - m_K^2 (-1 + 2 L_K) +\frac{5}{4} m_\eta^2  L_\eta 
- a_{6 \times \bar 6} \, \left(m_K^2 - m_\pi^2 \right)
\right] . 
\ \ \ 
\eea
\end{subequations}

Eq.~\eqref{eq:LEC1} implies that  the low-energy constants $a_{8\times8, 6 \times \bar 6}$  could be extracted  from 
lattice QCD  calculations  of $K^0$-$\bar{K}^0$ mixing  at different values of both $m_u=m_d$,   and $m_s$. 
Moreover,   at NLO one can derive  counterterm-free relations that connect ${\cal M}^{\pi \pi}$, 
${\cal M}^{K \bar K}$, and  $K \to \pi \pi$ matrix elements (${\cal  M}^{K \to \pi \pi}$). 
For the   ${\bf 8}_L \times {\bf 8}_R$ case, using the NLO Lagrangian of Ref.~\cite{Cirigliano:1999pv},  
we obtain 
\be
(2m_\pi^2-m_K\sq)F_\pi\sq {\cal M}^{\pi\pi}_{8 \times 8} = m_K\sq F_K\sq {\cal M}^{K\bar K}_{8 \times 8} 
\qquad  \qquad  \qquad \qquad \qquad \qquad \qquad \qquad \qquad \qquad \qquad \qquad 
\nn
\ee
\be
\qquad \qquad \qquad  
+ \frac{4iF_\pi\sq F_K}{\sqrt{2}}\bigg[2(m_\pi\sq-m_K\sq) {\cal M}^{K\to\pi^+\pi^-}_{8\times8} -(m_K\sq+2m_\pi\sq) {\cal M}_{8 \times 8}^{K\to \pi_0\pi_0}\bigg]
+ \Delta^{\rm loop}~,  \qquad
\label{CounterRelation}
\ee
where $\Delta^{\rm loop}$ is a calculable loop correction. 
In practice, at the moment neither of these two  approaches is feasible, due to missing or not sufficiently precise lattice input.
So in our estimates we adopt the following strategy:  
to obtain central values for $\Delta_{8\times8, 6 \times \bar 6}$, 
we evaluate the chiral loops at the scale $\mu_\chi = m_\rho$  and set the  counterterms to zero.
We then assess the counterterm  uncertainty in two ways: 
first,  assuming naive dimensional analysis  (NDA), namely $|a_{8\times8, 6 \times \bar 6}| \sim O(1)$,  we find  
$\Delta_{8 \times 8} = 0.02(20)$
and 
$\Delta_{6 \times \bar 6} = 0.07(20)$. 
Second,  requiring that the  counterterms  be of comparable size to their beta-functions, 
namely  $\Delta_n^{\rm (ct)} = \pm  | d \Delta_n^{\rm (loops)}  / d (\log \mu_\chi)  |$  
($n = 8\times8$ or $6 \times \bar 6$), 
we find  
$\Delta_{8 \times 8} = 0.02(36)$
and 
$\Delta_{6 \times \bar 6} = 0.07(16)$. 
To account for the strong scale dependence of loops in $\Delta_{8 \times 8}$, 
we enlarge the NDA estimate to  $\Delta_{8 \times 8}^{\rm (ct)} = \pm 0.3$ and 
use in  the subsequent analysis  $\Delta_{8 \times 8} = 0.02(30)$
and $\Delta_{6 \times \bar 6} = 0.07(20)$. 
%
%
The above results  point  to the fact that the dominant $SU(3)_L \times SU(3)_R$ correction to the relations 
\eqref{eq:relation1} is  captured by the ratio $(F_\pi/F_K)^2$ in \eqref{eq:relation2}.
Putting together the effect of  chiral loops and $F_K/ F_\pi = 1.19$~\cite{Aoki:2016frl},  
the total  chiral corrections in \eqref{eq:relation2} amount to 
$R_{8 \times 8} = 0.72(21)$  ($\sim 30 \%$ uncertainty) 
and $R_{6 \times \bar 6}= 0.76(14)$   ($\sim 20 \%$ uncertainty). 
%
Given that the chiral expansion is well behaved for these quantities,  we expect  
residual higher order corrections not to exceed 10\%,  well within the assigned ranges.

Using Eqs.~\eqref{eq:relation2} and the matrix elements of  the  $\Delta S =2$  operators 
calculated in Refs.~\cite{Carrasco:2015pra,Bertone:2012cu,Jang:2015sla,Boyle:2012qb,Garron:2016mva}
we find for the two-pion matrix elements renormalized 
in the $\overline{\rm MS}$ scheme  at the scale $\mu = 3$~GeV
\begin{subequations}
\bea
\langle \pi^+ |  O_2  | \pi^- \rangle  & = &  - \frac{5}{12}  \, B_2 \,  K \times R_{6 \times \bar 6}  
\qquad \qquad  \   K = \frac{2  \, F_K^2 \, m_K^4}{(m_d + m_s)^2}  
\qquad
\\
\langle \pi^+ |  O_3  | \pi^- \rangle  & = &     \frac{1}{12}  \, B_3 \,  K  \times R_{6 \times \bar 6}  
\\
\langle \pi^+ |  O_4  | \pi^- \rangle  & = &  - \frac{1}{3}  \, B_5 \,  K  \times R_{8 \times 8}  
\\
\langle \pi^+ |  O_5  | \pi^- \rangle  & = &  - B_4 \, K   \times R_{8 \times 8}  ~, 
\eea
\label{eq:results}%
\end{subequations}
where  the dimensionless scale- and scheme-dependent $B_{2,3,4,5}$ are  reviewed  in \cite{Aoki:2016frl}.~\footnote{Our operators $O_i$ are related to the $Q_i$ of Ref.~\cite{Aoki:2016frl} as follows: 
$O_{1,2,3} = (1/4) Q_{1,2,3}$, 
$O_{4,5} =  -(1/2)  Q_{5,4}$.} 
To  obtain the central value estimates for the matrix elements  we  use 
$F_K = 110$~MeV,  
$m_d (3 {\rm GeV}) =  4.3$~MeV,   
$m_s (3 {\rm GeV}) =  87.5$~MeV~\cite{Agashe:2014kda}. 
For the $B_i$ we take the midpoint of a conservative range encompassing the maximum 
and minimum values of the $N_f = 2+1$~\cite{Jang:2015sla},
 and $N_f = 2+1+1$~\cite{Carrasco:2015pra} 
  results  summarized  in Ref.~\cite{Aoki:2016frl}. 
The uncertainty associated with this treatment of the $B_{i}$ is at the level of 10\% for $B_{2,3}$,  20\% for $B_4$,  and 30\% for $B_5$. 
For the matrix elements of $O_{2,3}$ the uncertainty due to the quark masses is non negligible, but 
subdominant compared to the effect of NLO  chiral corrections. 
We summarize our current best estimates for the matrix elements and their uncertainties  in Table~\ref{tab:results}.
The  fractional uncertainty  is at the 20\%  level for  $\langle \pi^+ |  O_{2,3}  | \pi^- \rangle$  (dominated by chiral corrections), 
at the 35\% for $ \langle \pi^+ |  O_{5}  | \pi^- \rangle$  (dominated by chiral corrections), 
and at the 40\% level for $\langle \pi^+ |  O_{4}  | \pi^- \rangle$ 
(equally shared by chiral correction and lattice QCD input).

The effective coupling $g_{8 \times 8}$  can also be extracted  from the  electroweak penguin matrix 
elements $ \langle (\pi \pi)_{I=2} | {\cal Q}_{7,8} | K^0\rangle$~\cite{Blum:2012uk,Blum:2015ywa}.
This extraction was recently updated in Ref.~\cite{Cirigliano:2016yhc} to LO in the chiral expansion 
(in \cite{Cirigliano:2016yhc}  the notation $g_{8 \times 8}^{(i)} \to -  {\cal A}_{iLR}$ was used).   
Using the  value of $g_{8 \times 8}$  from  Ref.~\cite{Cirigliano:2016yhc} in 
Eq.~\eqref{eq:88pipi} and neglecting chiral corrections leads to  
$\langle \pi^+ |  O_4  | \pi^- \rangle  = - 1.9  \times 10^{-2}  \ {\rm GeV}^4$
and 
$\langle \pi^+ |  O_5  | \pi^- \rangle  = -  8.5  \times 10^{-2}  \ {\rm GeV}^4$, in  reasonable agreement 
with the  estimate of these matrix elements based on $K^0$-$\bar{K}^0$ mixing given in Eq.~\eqref{eq:results} and Table~\ref{tab:results}.
NLO chiral effects in $K^0 \to (\pi \pi)_{I=2}$  change the extracted  low-energy constant  as follows,  
 $g_{8 \times 8}  \to   g_{8 \times 8} /(1 + \Delta_2)$,  with 
 $\Delta_2 = - 0.30 \pm 0.20$~\cite{Cirigliano:2001hs}, where 
 the central value stems from chiral loop and known counterterms, 
 while the error encompasses an estimate of the unknown counterterms. 
Taking this into  account and  keeping the chiral logs in Eq.~\eqref{eq:88pipi} leads to 
  $\langle \pi^+ |  O_4  | \pi^- \rangle  = - 2.7  \times 10^{-2}  \ {\rm GeV}^4$
and 
$\langle \pi^+ |  O_5  | \pi^- \rangle  = -  12.7   \times 10^{-2}  \ {\rm GeV}^4$, 
in  excellent agreement  with the results of Table~\ref{tab:results}.

\begin{table}[t]
\begin{center}
\begin{tabular}{|ccc|}
\hline
\hline
$\langle \pi^+ |  O_1  | \pi^- \rangle $ & = &   $  (1.0  \pm 0.1  \pm 0.2)   \times 10^{-4}  \ {\rm GeV}^4 $
\\
$\langle \pi^+ |  O_2  | \pi^- \rangle $ & = &   $ - ( 2.7  \pm  0.3  \pm 0.5)  \times 10^{-2}  \ {\rm GeV}^4 $
\\
$\langle \pi^+ |  O_3  | \pi^- \rangle $ & = &     $  (0.9  \pm 0.1 \pm 0.2)  \times 10^{-2}  \ {\rm GeV}^4$
\\
$\langle \pi^+ |  O_4  | \pi^- \rangle $ & = &  $ - (2.6 \pm  0.8  \pm 0.8)  \times 10^{-2}  \ {\rm GeV}^4$
\\
$\langle \pi^+ |  O_5  | \pi^- \rangle  $ & = &  $-  (11  \pm  2  \pm  3)  \times 10^{-2}  \ {\rm GeV}^4$
\\
\hline
\hline
\end{tabular}
\end{center}
\caption{\small Pionic matrix elements of the operators in Eqs.~\eqref{eq:basis} in the $\overline{\rm MS}$ scheme at the scale $\mu = 3$~GeV.
The first uncertainty refers to the lattice QCD input on kaon matrix elements. 
The second uncertainty is associated to  the  size of  partially known NLO chiral corrections (only loops are taken into account)  and possible higher order effects. 
See text for discussion.}
\label{tab:results}  
\end{table}

{\bf  Determination of $\mathbf{\langle \pi^+ | O_{1}| \pi^- \rangle}$} --   
For completeness, we also update the analysis of Ref.~\cite{Savage:1998yh}. 
First, note that $O_1$  belongs to the ${\bf 27}_L \times {\bf 1}_R$ representation of $SU(3)_L \times SU(3)_R$, 
along with   $O_{\Delta S = 2}  =   \bar{s} \gamma_\mu (1 - \gamma_5) d \,  \bar{s} \gamma^\mu (1 - \gamma_5) d$ 
and the   component   $Q_2^{(27 \times 1)}$   of the $\Delta S=1$ operator  $Q_2  =   \bar{s} \gamma_\mu (1 - \gamma_5) u \,  \bar{u} \gamma^\mu (1 - \gamma_5) d = 
Q_2^{(27 \times 1)}  +  Q_2^{(8\times1)}$.
Using the normalization conventions of Ref.~\cite{Cirigliano:2003gt}, the leading order chiral realization of 
these operators is~\footnote{
Explicitly the projection reads 
$Q_2^{(27 \times 1)} =  2/5  \
 [\bar{s} \gamma_\mu (1 - \gamma_5) u \,  \bar{u} \gamma^\mu (1 - \gamma_5) d   + 
\bar{s} \gamma_\mu (1 - \gamma_5) d \,  \bar{u} \gamma^\mu (1 - \gamma_5) u)]  - 1/5  \
[ \bar{s} \gamma_\mu (1 - \gamma_5) d \,  \bar{d} \gamma^\mu (1 - \gamma_5) d   
+  \bar{s} \gamma_\mu (1 - \gamma_5) d \,  \bar{s} \gamma^\mu (1 - \gamma_5) s ] $.}
\begin{subequations}
\bea
Q_2^{(27 \times 1)} 
 & \to &   
 g_{27 \times 1}   \  {F_0^4}  \  \left( L_{\mu 32} L^\mu_{11}  + \frac{2}{3}  L_{\mu 31} L^\mu_{12} \right) 
 \\
 O_{\Delta S =2} 
 & \to &   
\frac{5}{3}  \,  g_{27 \times 1}   \  {F_0^4}  \   L_{\mu 32} L^\mu_{32}  
\\
4 \, O_1  
 & \to &   
\frac{5}{3}  \,  g_{27 \times 1}   \  {F_0^4}  \   L_{\mu 12} L^\mu_{12}  ~, 
\eea
\label{eq:chiral2}%
\end{subequations}
with  $L^\mu_{ij} = i  (U^\dagger  \partial^\mu U)_{ij}$. 
The factor of 4 multiplying $O_1$ in \eqref{eq:chiral2} accounts for the different normalization of $O_1$ 
compared to $Q_2$ and $O_{\Delta S= 2}$. 
In principle, Eqs.~\eqref{eq:chiral2}   allow one to  relate   $\langle \pi^+ | O_{1}| \pi^- \rangle$ to both $K^0$-$\bar{K}^0$ mixing (as done for 
 $\langle \pi^+ | O_{2,...,5}| \pi^- \rangle$) and  $\langle \pi^+ \pi^0 |Q_2| K^+\rangle$. 
However, it turns out that the determination in terms of  $K^0$-$\bar{K}^0$ mixing suffers from  potentially large  chiral corrections. 
The NLO analysis leads to 
\bea
\langle \pi^+ | O_{1}| \pi^- \rangle 
 &=& \frac{1}{4} \frac{m_\pi^2 F_\pi^2}{m_K^2 F_K^2}  \ 
 \langle \bar{K}^0 |  O_{\Delta S =2} | K^0 \rangle \ 
  \left( 1  + \Delta_{27\times 1}\right) 
\label{eq:O1fromBK}
\eea
with  a  strongly scale-dependent one-loop correction given  by 
$\Delta_{27\times 1}^{\rm (loops)}  =  \{ + 0.24,  -0.11, - 0.39 \}$ at $\mu_\chi = \{ m_\eta, m_\rho, 1~{\rm GeV} \}$, respectively. 
This is not unexpected, as  large chiral corrections to   $ \langle \bar{K}^0 |  O_{\Delta S =2} | K^0 \rangle$ were already found 
in Refs.~\cite{Bijnens:1984ec} and \cite{Becirevic:2004qd}. 
Taking for $B_K$ the midpoint of a range  that includes the $N_f = 2+1$ and $N_f = 2+1+1$ lattice results~\cite{Aoki:2016frl},  
namely    $B_K (3~{\rm GeV}) = 0.52 (3)$,   from \eqref{eq:O1fromBK} at $\mu_\chi = m_\rho$ we obtain 
$\langle \pi^+ |  O_1  | \pi^- \rangle = 0.97 \times 10^{-4}~{\rm GeV}^4$. 
The uncertainty from chiral corrections is at least 40-50\%, given the strong scale dependence of the one-loop effects. 
In light of this,  we focus next on the relation  of $\langle \pi^+ | O_{1}| \pi^- \rangle$ to $K^+ \to \pi^+ \pi^0$~\cite{Savage:1998yh}. 

At one loop in chiral perturbation theory we find 
\bea
\langle \pi^+ | O_{1}| \pi^- \rangle 
 &=& \frac{5}{3}  \  g_{27 \times 1} \  m_\pi^2 F_\pi^2 \left\{ 1 + \frac{m_\pi^2}{(4 \pi F_0)^2} (-1 + 3 L_\pi)   + \delta^{\pi \pi}_{27 \times 1} 
  \right\}.
\label{eq:o27pipi}
  \\
  \langle \pi^+ \pi^0  |  i Q_2 |  K^+ \rangle 
 &=& \frac{5}{3}  \  g_{27 \times 1} \  F_\pi  \ \left(m_K^2 - m_\pi^2 \right)   \ \left\{ 1 +  \Delta_{27}^{K^+  \pi^+ \pi^0}
  \right\}.
\label{eq:kpipi}
\eea
where  $\delta^{\pi \pi}_{27 \times 1}$  denotes  a linear combination of counterterms.   
$\Delta_{27}^{K^+ \pi^+ \pi^0}$ receives both loop and counterterms contributions:  
the loops have been computed in Ref~\cite{Cirigliano:2003gt}  and they are small while displaying a very mild scale dependence. 
The counterterms have been estimated in the large-$N_C$ approximation 
and are negligible~\cite{Cirigliano:2003gt}. Overall,  one finds $|1 + \Delta_{27}^{K^+ \pi^+ \pi^0} | = 0.98 \pm 0.05$~\cite{Cirigliano:2003gt}.
Comparing Eq.~\eqref{eq:kpipi} to the lattice QCD  results for 
$\langle \pi^+ \pi^0  | Q_2 |  K^+ \rangle$~\cite{Blum:2012uk,Blum:2015ywa} (in the $\overline{\rm MS}$ scheme at $\mu = 3$~GeV) 
we obtain  $g_{27 \times 1} =  0.34 (3)_{\rm LQCD}  (2)_\chi$,  where the first error is from the lattice input and the second from the 
chiral corrections in Eq.~\eqref{eq:kpipi}~\footnote{This result is in good agreement with  $g_{27} \simeq 0.29$  found 
by a fit to the $K \to \pi \pi$ decay rates~\cite{Cirigliano:2003gt}.   The identification $g_{27} =  g_{27 \times 1}$, however, neglects 
mixing of $Q_2$ with other operators and its scale dependence.  
Our result is also in good agreement with Ref~\cite{Savage:1998yh}, once we take into account that  
the coupling $g^{(\bf{27})}$ of Ref.~\cite{Savage:1998yh} is related to our 
$g_{27 \times1}$ by  $g^{(\bf{27})} = (5/12)  g_{27\times1}$. }.   
Using this value in Eq.~\eqref{eq:o27pipi}, and assigning a conservative 20\% error 
due to the unknown counterterms  in $\delta^{\pi \pi}_{27 \times 1}$, we  obtain the   result reported in Table~\ref{tab:results}.
Finally, note that the determination of  $\langle \pi^+ | O_{1}| \pi^- \rangle$ from $K^0$-$\bar{K}^0$ mixing, 
though plagued by larger uncertainty,    is quite consistent  with the result of Table~\ref{tab:results}.

 {\bf  Discussion and conclusion} -- 
In this letter we  have provided estimates for the $\pi^- \to \pi^+ $ matrix elements 
of  all  Lorentz scalar $\Delta I = 2$  four-quark operators relevant to the study of 
TeV-scale lepton number violation.    The  analysis is based on 
(i)  chiral $SU(3)$ symmetry, which relates the $\pi^- \to \pi^+$  matrix elements of 
$O_{1,...,5}$ defined in  Eq.~\eqref{eq:basis} to the  $K^0 \to \bar{K}^0$  and  $K \to \pi \pi$  
matrix elements of their $\Delta S = 2$ and $\Delta S = 1$ chiral partners;
(ii)  lattice QCD input for the relevant kaon matrix elements. 
Our main results are summarized in   Eqs.~\eqref{eq:results} and Table~\ref{tab:results}.

A preliminary lattice QCD calculation of  the  matrix elements considered in this letter  
has appeared in Ref~\cite{Nicholson:2016byl}~\footnote{There is a slight difference between 
the operators used here  ($O_j$)  used  here and  the ones  used in Ref.~\cite{Nicholson:2016byl}
(${\cal O}^{++}_{i +}$).   Using parity-invariance of QCD, 
for the $\pi^- \to \pi^+$ matrix elements we have the following relations: 
$ \langle {\cal O}^{++}_{1+} \rangle =  \langle O_4 \rangle$, 
$ \langle {\cal O}^{'++}_{1+} \rangle =  \langle O_5 \rangle$, 
$ \langle {\cal O}^{++}_{2+} \rangle =   2 \  \langle O_2 \rangle$, 
$ \langle {\cal O}^{'++}_{2+} \rangle =   2 \  \langle O_3 \rangle$, 
$ \langle {\cal O}^{++}_{3+} \rangle =   2 \  \langle O_1 \rangle$.}.
A complete comparison is not yet possible  because Ref.~\cite{Nicholson:2016byl} presents  results  for  bare matrix elements. 
Nonetheless, already at this level,  we find the  hierarchy of bare matrix elements in \cite{Nicholson:2016byl} 
to be in qualitative agreement with our results. 

For all the symmetry relations used here,  we have included the NLO chiral  loop corrections, 
showing that the chiral expansion is well behaved and the relations are robust at the 20-30\% level, 
depending on the operator under consideration.  
The remaining uncertainty  can be further reduced as the precision on $K^0$-$\bar{K}^0$ matrix elements improves.
Our results provide a first controlled estimate of the hadronic matrix elements 
needed to  assess the sensitivity of  \NLDBD \ to TeV-scale sources of  lepton number violation, and 
can be used as input in  nuclear structure calculations of the leading pion-exchange operators~\cite{Engel:2016xgb}.

\newpage

\section*{Acknowledgements}
VC, MG and  EM acknowledge support by the LDRD program at Los Alamos National Laboratory.
WD  acknowledges  support by the Dutch Organization for Scientific Research (NWO) 
through a RUBICON  grant. 
We thank Brian Tiburzi for discussions on the operator basis.

\bibliographystyle{h-physrev3} 
\bibliography{bibliography}

\begin{thebibliography}{10}

\bibitem{Schechter:1981bd}
J.~Schechter and J.~W.~F. Valle,
\newblock Phys. Rev. {\bf D25}, 2951 (1982).

\bibitem{Davidson:2008bu}
S.~Davidson, E.~Nardi, and Y.~Nir,
\newblock Phys. Rept. {\bf 466}, 105 (2008), 0802.2962.

\bibitem{KamLAND-Zen:2016pfg}
KamLAND-Zen, A.~Gando {\em et~al.},
\newblock Phys. Rev. Lett. {\bf 117}, 082503 (2016), 1605.02889,
\newblock [Addendum: Phys. Rev. Lett.117,no.10,109903(2016)].

\bibitem{Alfonso:2015wka}
CUORE, K.~Alfonso {\em et~al.},
\newblock Phys. Rev. Lett. {\bf 115}, 102502 (2015), 1504.02454.

\bibitem{Albert:2014awa}
EXO-200, J.~B. Albert {\em et~al.},
\newblock Nature {\bf 510}, 229 (2014), 1402.6956.

\bibitem{Agostini:2013mzu}
GERDA, M.~Agostini {\em et~al.},
\newblock Phys. Rev. Lett. {\bf 111}, 122503 (2013), 1307.4720.

\bibitem{Gando:2012zm}
KamLAND-Zen, A.~Gando {\em et~al.},
\newblock Phys. Rev. Lett. {\bf 110}, 062502 (2013), 1211.3863.

\bibitem{Elliott:2016ble}
S.~R. Elliott {\em et~al.},
\newblock {Initial Results from the MAJORANA DEMONSTRATOR},
\newblock 2016, 1610.01210.

\bibitem{Andringa:2015tza}
SNO+, S.~Andringa {\em et~al.},
\newblock Adv. High Energy Phys. {\bf 2016}, 6194250 (2016), 1508.05759.

\bibitem{Weinberg:1979sa}
S.~Weinberg,
\newblock Phys. Rev. Lett. {\bf 43}, 1566 (1979).

\bibitem{Rodejohann:2011mu}
W.~Rodejohann,
\newblock Int. J. Mod. Phys. {\bf E20}, 1833 (2011), 1106.1334.

\bibitem{deGouvea:2013zba}
A.~de~Gouv{\^e}a and P.~Vogel,
\newblock Prog. Part. Nucl. Phys. {\bf 71}, 75 (2013), 1303.4097.

\bibitem{DellOro:2016tmg}
S.~Dell'Oro, S.~Marcocci, M.~Viel, and F.~Vissani,
\newblock Adv. High Energy Phys. {\bf 2016}, 2162659 (2016), 1601.07512.

\bibitem{Engel:2016xgb}
J.~Engel and J.~Men\'endez,
\newblock (2016), 1610.06548.

\bibitem{Pas:2000vn}
H.~P{\"a}s, M.~Hirsch, H.~V. Klapdor-Kleingrothaus, and S.~G. Kovalenko,
\newblock Phys. Lett. {\bf B498}, 35 (2001), hep-ph/0008182.

\bibitem{Prezeau:2003xn}
G.~Pr\'ezeau, M.~Ramsey-Musolf, and P.~Vogel,
\newblock Phys. Rev. {\bf D68}, 034016 (2003), hep-ph/0303205.

\bibitem{Graesser:2016bpz}
M.~L. Graesser,
\newblock (2016), 1606.04549.

\bibitem{Faessler:1996ph}
A.~Faessler, S.~Kovalenko, F.~Simkovic, and J.~Schwieger,
\newblock Phys. Rev. Lett. {\bf 78}, 183 (1997), hep-ph/9612357.

\bibitem{Savage:1998yh}
M.~J. Savage,
\newblock Phys. Rev. {\bf C59}, 2293 (1999), nucl-th/9811087.

\bibitem{Carrasco:2015pra}
ETM, N.~Carrasco {\em et~al.},
\newblock Phys. Rev. {\bf D92}, 034516 (2015), 1505.06639.

\bibitem{Bertone:2012cu}
ETM, V.~Bertone {\em et~al.},
\newblock JHEP {\bf 03}, 089 (2013), 1207.1287,
\newblock [Erratum: JHEP07,143(2013)].

\bibitem{Jang:2015sla}
SWME, B.~J. Choi {\em et~al.},
\newblock Phys. Rev. {\bf D93}, 014511 (2016), 1509.00592.

\bibitem{Boyle:2012qb}
RBC, UKQCD, P.~A. Boyle, N.~Garron, and R.~J. Hudspith,
\newblock Phys. Rev. {\bf D86}, 054028 (2012), 1206.5737.

\bibitem{Garron:2016mva}
RBC/UKQCD, N.~Garron, R.~J. Hudspith, and A.~T. Lytle,
\newblock JHEP {\bf 11}, 001 (2016), 1609.03334.

\bibitem{Gonzalez:2015ady}
M.~Gonz\'alez, M.~Hirsch, and S.~G. Kovalenko,
\newblock Phys. Rev. {\bf D93}, 013017 (2016), 1511.03945.

\bibitem{Gabbiani:1996hi}
F.~Gabbiani, E.~Gabrielli, A.~Masiero, and L.~Silvestrini,
\newblock Nucl. Phys. {\bf B477}, 321 (1996), hep-ph/9604387.

\bibitem{Buras:2000if}
A.~J. Buras, M.~Misiak, and J.~Urban,
\newblock Nucl. Phys. {\bf B586}, 397 (2000), hep-ph/0005183.

\bibitem{Donoghue:1982cq}
J.~F. Donoghue, E.~Golowich, and B.~R. Holstein,
\newblock Phys. Lett. {\bf B119}, 412 (1982).

\bibitem{Bijnens:1984ec}
J.~Bijnens, H.~Sonoda, and M.~B. Wise,
\newblock Phys. Rev. Lett. {\bf 53}, 2367 (1984).

\bibitem{Gasser:1984gg}
J.~Gasser and H.~Leutwyler,
\newblock Nucl. Phys. {\bf B250}, 465 (1985).

\bibitem{Cirigliano:2003gt}
V.~Cirigliano, G.~Ecker, H.~Neufeld, and A.~Pich,
\newblock Eur. Phys. J. {\bf C33}, 369 (2004), hep-ph/0310351.

\bibitem{Cirigliano:1999pv}
V.~Cirigliano and E.~Golowich,
\newblock Phys. Lett. {\bf B475}, 351 (2000), hep-ph/9912513.

\bibitem{Becirevic:2004qd}
D.~Be\'cirevi\'c and G.~Villadoro,
\newblock Phys. Rev. {\bf D70}, 094036 (2004), hep-lat/0408029.

\bibitem{Aoki:2016frl}
S.~Aoki {\em et~al.},
\newblock (2016), 1607.00299.

\bibitem{Agashe:2014kda}
Particle Data Group, K.~A. Olive {\em et~al.},
\newblock Chin. Phys. {\bf C38}, 090001 (2014).

\bibitem{Blum:2012uk}
T.~Blum {\em et~al.},
\newblock Phys. Rev. D {\bf 86}, 074513 (2012), 1206.5142.

\bibitem{Blum:2015ywa}
T.~Blum {\em et~al.},
\newblock Phys. Rev. D {\bf 91}, 074502 (2015), 1502.00263.

\bibitem{Cirigliano:2016yhc}
V.~Cirigliano, W.~Dekens, J.~de~Vries, and E.~Mereghetti,
\newblock (2016), 1612.03914.

\bibitem{Cirigliano:2001hs}
V.~Cirigliano and E.~Golowich,
\newblock Phys. Rev. {\bf D65}, 054014 (2002), hep-ph/0109265.

\bibitem{Nicholson:2016byl}
A.~Nicholson {\em et~al.},
\newblock {Neutrinoless double beta decay from lattice QCD},
\newblock in {\em {Proceedings, 34th International Symposium on Lattice Field
  Theory (Lattice 2016): Southampton, UK, July 24-30, 2016}}, 2016, 1608.04793.

\end{thebibliography}

\end{document}